- 

# High-Precision Real-Time Pores Detection in LPBF Using Thermal Energy Density (TED) Signals

Chuxiao Meng[1,2], Conor Porter[1], Sina Malakpour[1], Garrett Mathesen[1],

Seongyeon Yang[1,3], Jian Cao*[1]

1. Department of Mechanical Engineering, Northwestern University
2. Department of Material Science and Engineering, Northwestern University
3. Department of Public Policy, University of Chicago

# Abstract

**(Real time, high accuracy, low cost, scalable**

Pore formation during Laser Powder Bed Fusion (LPBF) has long posed challenges in metal 3D printing, significantly affecting the mechanical properties of the final product. Porosity frequently occurs because of an unstable keyhole formation, triggered by an excess laser energy. Traditional approaches for detecting pores rely heavily on CT scanning, a time-consuming and costly method unsuitable for large-scale production. In response to these limitations, we have developed a real-time pore detection method using thermal sensor data, offering a more efficient, cost-effective alternative for quality control during the LPBF process.

Our method, validated against CT-scanned pore counts, provides a high degree of accuracy, achieving an R² value of 0.94 between the across eight sample prints. This approach also effectively tracks pore formation trends as the layer-wise printing pattern changes, providing timely insights into product quality, which may serve as important datapoints for real-time adaptive parameters optimization in the future. In contrast to prior machine learning-based techniques, which were limited by high computational costs and lacked direct validation strategy, the method introduced in this paper is not only faster and simpler but also highly scalable for industrial use.

# 1.Introduction

Additive manufacturing, widely known as 3D printing, has transformed the production landscape by enabling the creation of complex and customized components directly from digital designs. Among the advanced techniques in this domain, Laser-Powder Bed Fusion (LPBF) has emerged as a leading method for fabricating intricate metallic parts with exceptional precision.[1][2][3] LPBF works by directing a high-energy laser beam onto a bed of fine metal powder, selectively melting and fusing the material layer by layer to build three-dimensional objects. This technology has unlocked new possibilities across industries—from aerospace to medical devices—by allowing for designs that are not feasible with traditional manufacturing methods. Nonetheless, challenges such as quality control and defect management, particularly the formation of porosity, continue to be areas of active research to ensure the reliability and mechanical integrity of LPBF-produced components.

The extreme thermal gradients and rapid heating and cooling cycles inherent in the Laser-Powder Bed Fusion (LPBF) process induce complex structural dynamics within the molten material. A prominent defect that arises under these conditions is keyhole porosity. This defect manifests when the laser's energy input is so high that it causes excessive vaporization of the metal powder, generating significant recoil pressure. This pressure pushes the surface of the melt pool downward, forming a deep and narrow cavity known as a keyhole. While the keyhole enhances laser energy absorption and improves processing efficiency, it also creates an imbalance among various forces at play—including recoil pressure, vapor flow dynamics, surface tension, and thermocapillary convection. [4][5][6][7]

This imbalance leads to unstable melt pool conditions, making the keyhole susceptible to collapse. When the keyhole collapses, the walls of the cavity rapidly close in, trapping vapor and forming gas bubbles within the melting pool. These bubbles can detach from the keyhole walls and become trapped as the material solidifies, resulting in pores embedded within the final structure.[4][8] Such porosity compromises the mechanical integrity of the fabricated component, potentially leading to reduced strength, diminished fatigue life, and overall weakening of the material. Detecting and controlling the keyhole collapse and subsequent pore formation during LPBF are crucial for improving the quality and reliability of parts produced by this process.

 Traditional methods, such as X-ray Computed Tomography (CT) scanning, have been widely used to detect these pores. However, CT is too slow and costly, making it impractical for large-scale industrial use. Recognizing this limitation, several recent studies have explored alternative approaches. Zhao et al. (2017) employed high-speed X-ray imaging to observe melt pool dynamics, enabling real-time monitoring of the LPBF process[4]. Ren et al. (2023) took a different approach by using convolutional neural network to track keyhole porosity generation in real time, offering a solution for detecting pore formation based on melt pool instability [9]. Malakpour Estalaki et al. (2022) developed machine learning models that utilize in-situ thermal imaging data to predict micropores during LPBF, improving defect detection by incorporating thermal history and neighboring voxel data[10]. However, these methods often rely on highly specialized equipment that is difficult to scale for industrial applications. Additionally, they lack direct comparisons with CT-scanned pore counts, limiting their ability to assess the accuracy of pore predictions across an entire 3D sample----the interaction of the laser with previously formed pores in earlier layers can complicate predictions, as pores may be moved, altered, or even healed during subsequent layers.

In response to these challenges, we developed a low-cost, real-time method for detecting pores using Thermal Energy Density (TED) signals during LPBF. This approach leverages the strong correlation between TED signal peaks and keyhole collapse events, which are precursors to pore formation. Unlike previous methods that depend on complex external imaging systems, our TED-based technique integrates sensor data directly from the manufacturing process, offering a more scalable and practical solution. To the best of our knowledge, this is the first time a pores detection method approximates the exact numbers of pores in an entire printed part and validated by the CT scanned pores count.

The experimental setup, conducted in collaboration with the Argonne National Laboratory Advanced Photon Source (APS) and Northwestern University, records real-time TED signals during the printing process [11]. By identifying TED "peaks" using a specially design algorithm, we were able to accurately identify keyhole collapse patterns and pores, achieving an $R^2 > 0.94$ correlation between TED peaks count and pores count, suggesting the pores count could be accurately estimated by tracking the regression fitting line. This direct comparison with CT-scanned pore counts validates the accuracy of our method. Moreover, the model effectively captured changes in layer-wise pores distribution when there is a change in printing pattern, demonstrating the robustness of the TED method and its potential for wide-scale industrial adoption.

# 2.Mehology

Fred et al.(2022)[11] shows that a keyhole collapse, which is a major cause of pores during LPBF, corresponds to Peaks in TED signal. In this study, the experiments are performed using the same integrated experimental setup. Then an algorithm was developed to identify peaks in Thermal Energy Density (TED) signals, measured by the heat sensor, during the L-PBF process to detect keyhole collapse and pores.

## 2.1 Experiment Set Up

### 2.1.1 The Melt Pool Monitoring System

The experiments were conducted in collaboration with the Argonne National Laboratory Advanced Photon Source (APS) and Northwestern University. A coaxial photodiode system, same as the one introduced in Fred et al. (2022)[11], was integrated into the L-PBF process to monitor real-time thermal emissions from the melt pool. A schematic of the apparatus is shown in Fig. xx. The system operated at a sampling rate of 200 kHz, allowing for high-resolution detection of thermal energy fluctuations. The TED signal was derived from a wide bandwidth range of 400-900 nm, with two additional narrow bandwidth channels centered at 680 nm and 700 nm for further spectral analysis. The output from the wide bandwidth channel provided the TED metric, which is crucial for tracking energy input and identifying melt pool instabilities.

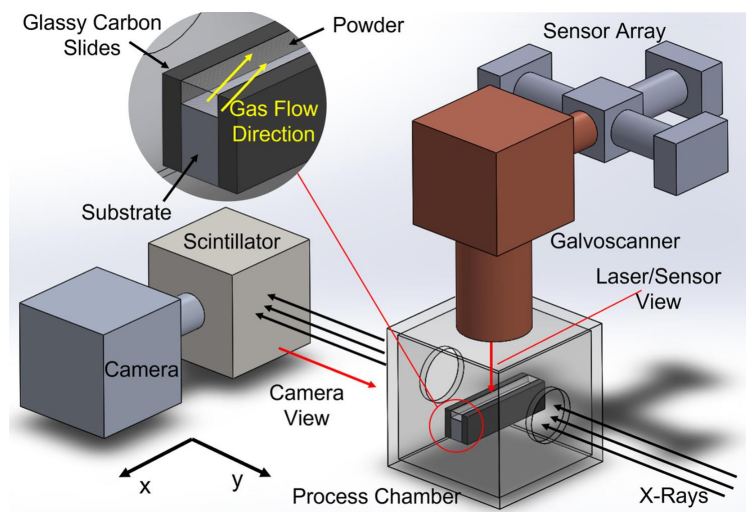

Fig. xx. Setup of the melting pool monitoring system

This setup was calibrated using a tungsten bulb, covering a temperature range from 1400 °C to 2500 °C, to ensure accurate readings throughout the experiment. The calibration curve exhibited an R² value of 0.998, confirming the reliability of the TED sensor for this application.

Two sets of L-PBF experiments with the set up mentioned above were conducted: the Advanced Photon Source (APS) experiments and the PART experiment. In APS experiments, the laser only scans one line, or two lines with a turn-around; the high-speed X-ray was applied to monitor the real-time melting pool shape. In PART experiment, eight samples with the same geometry but different processing parameters were printed. The TED values were recorded through the entire manufacturing time, but the printing processes were not monitored by the high-speed X-Ray camera.

### 2.1.2 Advanced Photon Source (APS) Experiment

The APS experiments utilized SLM-manufactured samples as substrates. The substrates were fabricated using the AlSi10Mg-63/20-AMS powder on a DMG MORI LASERTEC 12 SLM system, with parameters such as 240W laser power, 1550mm/s speed, and 150μm hatch spacing. These substrates were printed as 400μm thick, 50mm long walls, machined to 3mm in height, and manually ground to ensure flatness. Glassy carbon slides were used to sandwich the samples, and the powder layer thickness was determined by the height difference between the substrate and slides. A folded track scanning strategy with 0.1mm toolpath spacing and a 500mm/s scan speed was employed. The laser used was an nLight AFX 1000 with a Gaussian profile.

Table 1. Detailed list of experiments analyzed

| Exp. | Laser Power | Scan Speed | Powder layer | Gas Flow | Scan Pattern |
|---|---|---|---|---|---|
| A | 200 W | 500mm/s | 62 μm | 0 LPM | 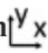 |
| B | 200 W | 500mm/s | 95 μm | 2.8 LPM | 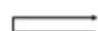 |
| C | 200 W | 500mm/s | 102 μm | 2.8 LPM | 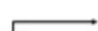 |
| D | 200 W | 500mm/s | 183 μm | 0 LPM | 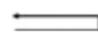 |

### 2.1.3 The Whole Sample Printing Experiment

Eight whole part samples with identical geometry were printed using varying processing parameters. The geometry of the samples, as illustrated in Fig xx, consists of multiple characteristic geometric features. The lower part provides a stable bulk foundation, while the top part includes cylindrical and rectangular features of varying dimensions. Each part was printed on a LASERTEC 12 SLM machine (DMG MORI, Bielefeld, Germany) with AlSi10Mg-63/20-AMS powder (TEKNA, Montreal, CA).

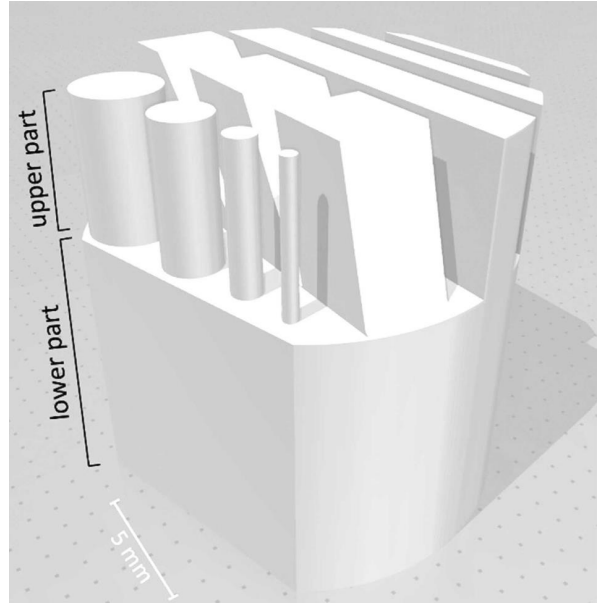

Fig xx. The geometry of the samples that were used in this study.

The processing parameters varied across the eight samples, with adjustments made to both hatch speed and hatch power. These parameters were critical in determining the energy input into the material. Table 2 shows the specific values used in the experiment, where the hatch speed ranged from 900 mm/s to 1800 mm/s, and the hatch power ranged from 120 W to 360 W.

Table 2. The processing parameter list of samples in PART experiment

|                    | Sample 1 | Sample 2 | Sample 3 | Sample 4 | Sample 5 | Sample 6 | Sample 7 | Sample 8 |
|--------------------|----------|----------|----------|----------|----------|----------|----------|----------|
| Hatch Speed (mm/s) | 1800     | 1550     | 1350     | 1200     | 1200     | 1550     | 1200     | 900      |
| Hatch Power (W)    | 360      | 360      | 360      | 360      | 240      | 240      | 120      | 360      |

The TED (Thermal Energy Density) values were continuously recorded throughout the printing process for each sample. TED data was mapped to the x, y coordinates and layer numbers, allowing for precise tracking of the thermal history at both the bulk and top parts of the samples.

## 2.2 An Overview of TED Signals during LPBF

Upon initial examination, the TED signals, as illustrated in Figure xxx, appear to exhibit random and disordered fluctuations. These signals were obtained from the PART samples during the LPBF process.

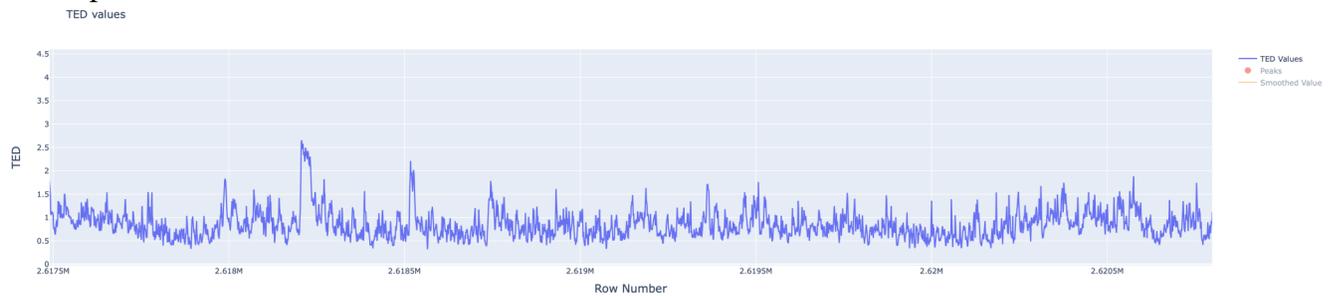

Figure xxx. Example of TED signals distribution for PART samples.

Fred et al. (2022) have demonstrated a strong correlation between TED peaks and keyhole collapse in the APS experiments. However, several challenges complicate the accurate identification of these peaks among millions of TED values in the whole part. These challenges are outlined below:

1. Signal sensitivity: TED signals exhibit different sensitivities between APS experiments and the PART samples. In the PART samples, the magnitude of signal fluctuations is generally larger than those observed in APS experiments, rendering the peaks less prominent.
2. Fluctuations: The TED signals inherently display general fluctuations, which may obscure the more significant peak events related to keyhole collapse.
3. Data variability: TED signals exhibit diverse patterns across millions of recorded data points, further complicating the process of identifying consistent peak structures.

## 2.3 TED Peaks Identification Method

### 2.3.1 Approximation of the TED distribution using a third-degree polynomial.

As aforementioned, TED signals have general fluctuations due to several reasons, such as the fluctuation of power and error of the sensor. Such fluctuations may bring "noise" for any peaks identifying strategy. To mitigate the effect of such "noise," we first "smooth" the TED signals

using a cubic polynomial. A cubic polynomial that approximates the data in each window can be expressed as:

$$P(x) = a_3 x^3 + a_2 x^2 + a_1 x + a_0$$

where $a_3$, $a_2$, $a_1$, $a_0$ are the polynomial coefficients, and x represents the position (index) of a data point within the window. For each window of n data points (here, n=600), the Savitzky-Golay filter was applied to fits a cubic polynomial to minimize the squared error between the actual data values and the polynomial's estimated values. The TED values for the window are denoted as $y_1, y_2, \ldots, y_n$, and the corresponding positions (indices) are $x_1, x_2, \ldots, x_n$. The smoothed value was obtained by identifying coefficients $a_3$, $a_2$, $a_1$, $a_0$ that minimize the sum of squared residuals:

$$\min \sum_{i=1}^{n} (y_i - P(x_i))^2$$

Figure xxx shows the smoothed values of TED signals, showing in orange curve. By doing this, we can identify TED peak reference to the "smoothed values" instead of the raw signal, thus reducing noise and errors.

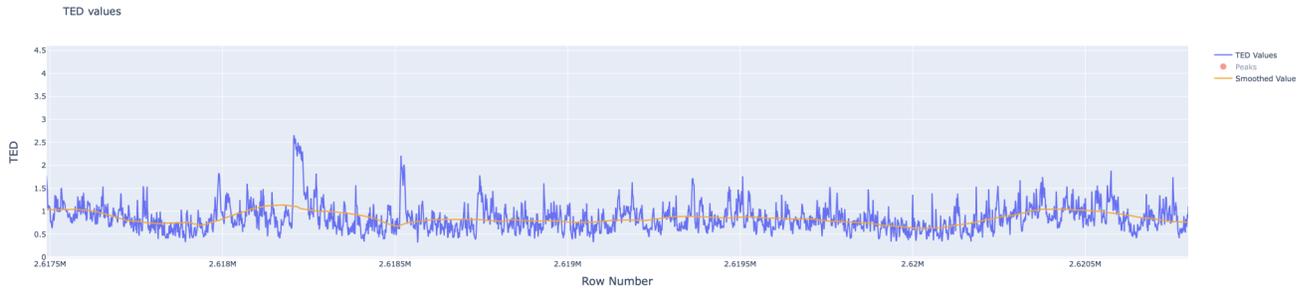

Figure xxx. TED signals distribution with smoothed values. The orange curve shows the smoothed curve.

## 2.3.2 Identifying TED patterns that correspond to keyhole collapse

For most of the printing, the TED values fluctuate up and down in a small time interval. During a keyhole collapse, the TED signals first rise to a significantly higher value, then drop to the normal value—a "peak" forms. Pores formed when there is a keyhole collapse.

After closely examining the identified TED peaks in the APS data, we found that during keyhole collapse, particularly in the expansion and collapse stages, TED values remain elevated for more than 0.07 ms (approximately 15 rows in the dataset). In contrast, during other scenarios, the signal fluctuates within approximately 0.02 ms (around 4 rows). This means that in rows where TED peaks occur, more than 15 consecutive rows show higher magnitudes compared to their neighbors. We utilized this feature to develop a peak identification algorithm.

Building on the signal smoothing step and the observed TED patterns corresponding to keyhole collapse, a TED peak identification algorithm was designed to detect keyhole collapses and predict pore formation. The algorithm works as follows: first, for each row, the difference between the original TED values and the smoothed values was computed to highlight deviations. An interval was classified as a peak if it contains more than a specified number of consecutive positive deviations (adjustable) with difference values greater than either [H (adjustable, default value = 0.335)] or [smoothed value × M (adjustable, default value = 0.2)].

The adjustable parameters (H, M) represent the absolute deviation (H) and relative deviation (M) of the TED values from the smoothed values. The default relative deviation, M=0.2, was determined by analyzing the general distribution of TED values and TED peaks, while the absolute deviation, H=0.335, was chosen to handle fluctuations in the signals: for signals with locally higher distributions, relying solely on relative deviation may cause peaks to be missed. Fig. xxx illustrates the labeled TED peaks for a randomly selected printing interval.

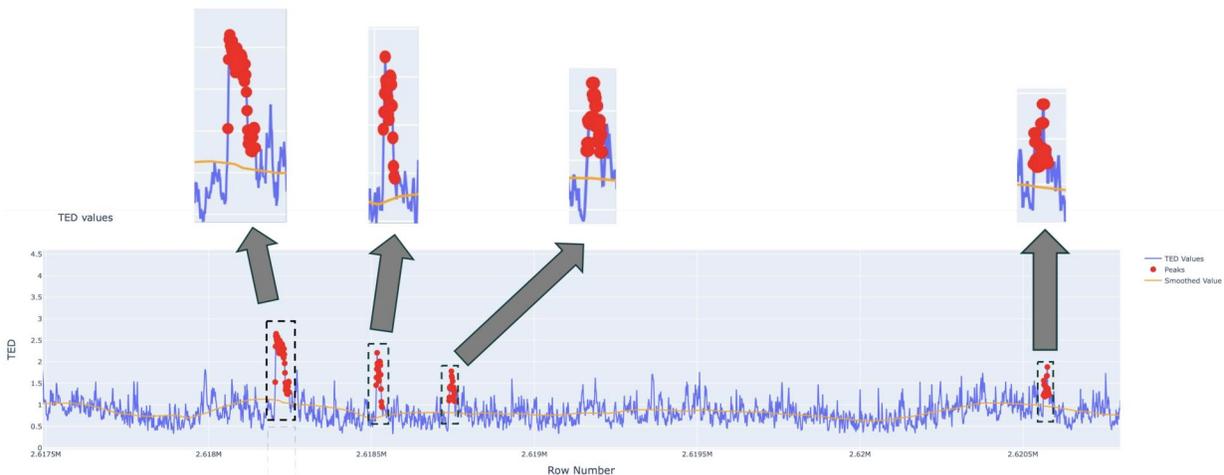

Fig xxx. Illustration of the TED peaks identification method. The orange line is the third-degree polynomial that approximates the distribution of TED values; the red dots are labeled TED peaks intervals, corresponding to keyhole collapses.

## 2.4 TED Peaks Count and Pores Count across Eight Samples

For each PART sample, the TED peak, which consist of consecutive TED datapoints with varying sized, were identified. Now we want to count the number of such peaks for each sample. The row numbers of peak intervals (at least 15 consecutive datapoint) identified by the model were recorded in a new list, thus we have the list of all peak intervals. Then, an algorithm counts the number of consecutive peak sequences in a list of peak indices by iterating through the list and identifying groups of consecutive values. It initializes a counter for these sequences and uses a nested loop to detect when consecutive peaks occur (i.e., when one peak's index is immediately followed by the next). As it identifies each sequence of consecutive peaks, it increments the counter. The process continues until all indices in the list have been checked, and the final count of consecutive sequences, representing the number of peaks, is recorded.

The exact pores count in eight samples were measured by CT scanning to compare with TED peaks identified in each sample. The CT scanning was performed using a Zeiss Metrotom 800 HR CT system, equipped with a minimum focal spot size of 7 μm and an accelerating voltage range of 30-225 kV at a 0-3000 μA tube current. The scanning parameters were set to use a 160 kV X-ray source, a 121 μA current, a 2 mm aluminum filter, and an exposure time of 1 second, with 3 frames averaging and 580 projections/rotation. The voxel resolution was 17.3 μm, and the scanned images were reconstructed using FDK reconstruction methods. Fig. xx shows the example of visualizations of the defects detected by CT.

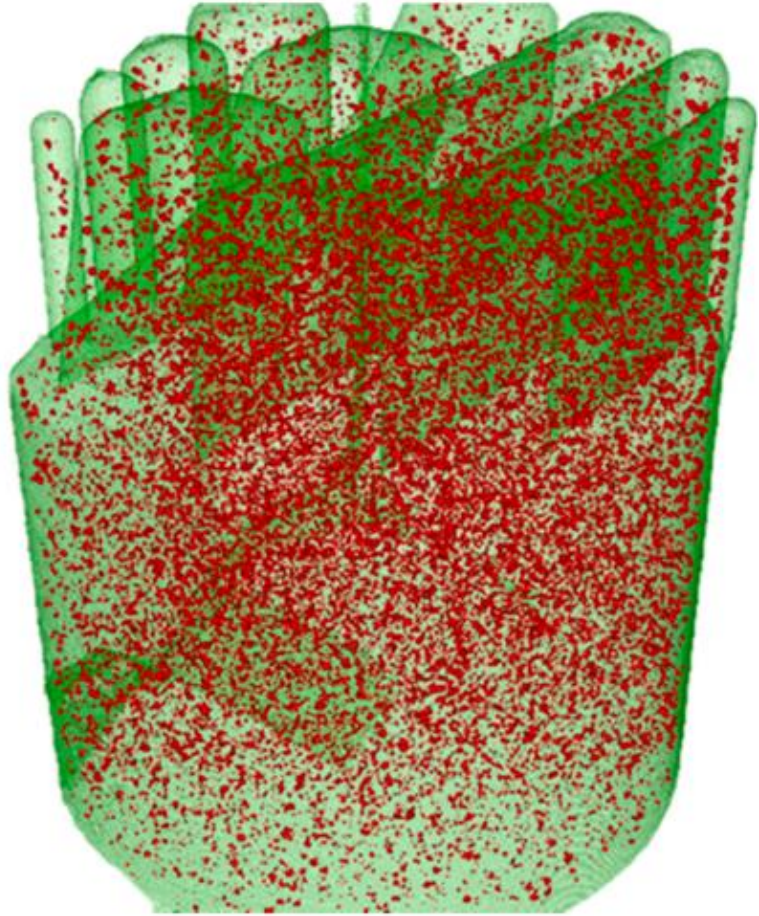

Fig xx. Pores (red points) detected by CT method.

## 2.5 Evaluating Layer-Wise Performance of the Model

The layer-by-layer process of Additive Manufacturing (AM) helps analyze how changes in the printing pattern affect pores across layers. In this study, each sample was printed with 573 layers, each having a thickness of 30 μm.

 The CT scanning process captured 2500 layers with a resolution of 10 μm per layer. However, these 2500 CT layers do not directly correspond to the 573 printed layers. The CT scan began from the bottom of the sample, which included the support structure beneath the printed part. As the scan reached the top, it continued past the sample, capturing part of the air above the sample as well. This extended scanning area provided additional context but did not affect the analysis of the printed structure itself.

The 573 printed layers correspond approximately to the CT-scanned layers 300-2000, with particular attention given to the transition around layer 400. At this point, the geometry shifted from a bulk cylindrical lower section to a more complex top part composed of rectangular and circular features. In the bulk part (before layer 400), the larger printed area exhibited a relatively simpler laser-scanning pattern, reducing the likelihood of pore formation. In contrast, the top part (after layer 400) involved more intricate scanning patterns, including turns and changes in direction, which resulted in a higher density of pores. Fig .xx illustrates the printing patterns and the TED distribution before and after layer 400**.**

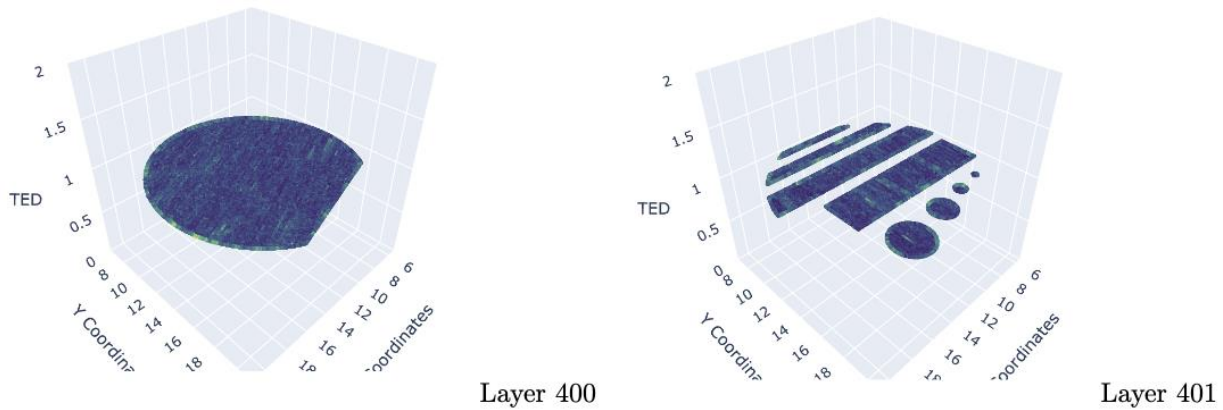

Layer 400　　　　　　　　　　　　　　　　　　　　　　Layer 401

Figure xxx. Printing pattern of samples before and after 400 layers.

## 2.6 X-Ray Ground Truth for the peak identified by the model

Fred et al.(2022) use an automatic deep learning segmentation model based on U-Net with a VGG-16 encoder employed. The U-Net output images, as shown in Fig. xx, introduced an typical keyhole collapse cycle. Initially stable (t = 3.50 ms), the keyhole rapidly enlarges and distorts (t = 3.52 to 3.62 ms), collapsing between t = 3.66 and 3.74 ms. Afterward, a stable keyhole re-forms (t = 3.78 ms), with characteristic oscillations of the melt pool surface starting at t = 3.58 ms. This suggests that a typical keyhole collapse includes 4 stages: stable>>expansion>>collapse>> recovery.

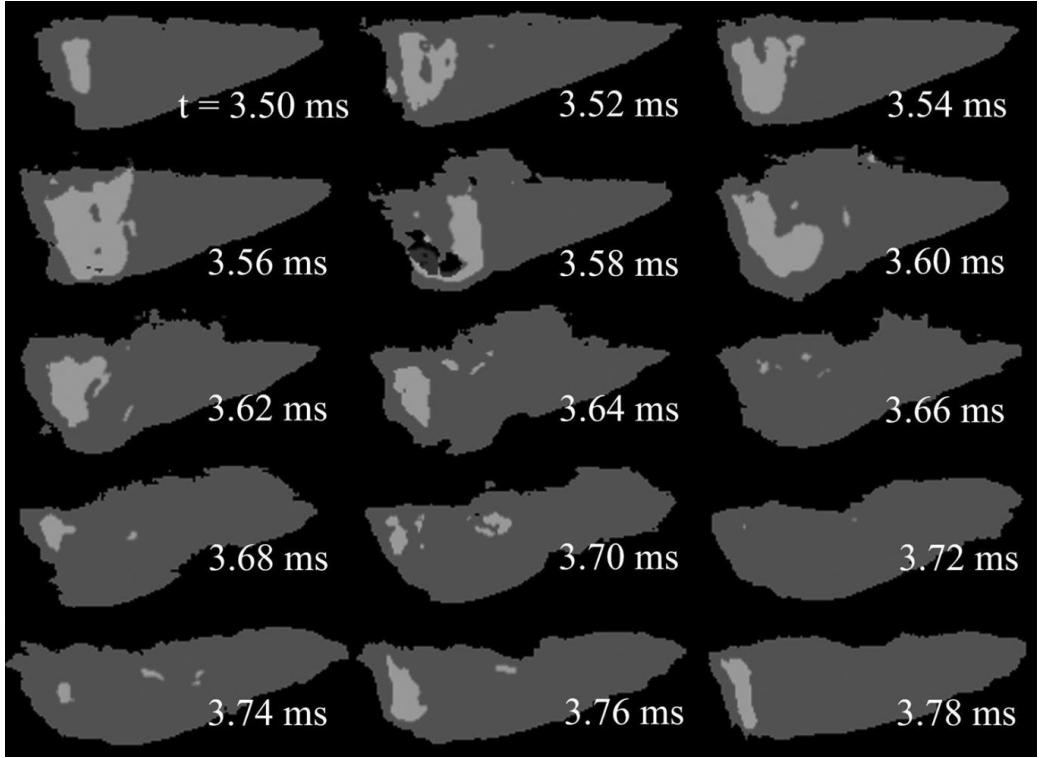

Fig.xx. Melt pool (dark gray) and keyhole/vapor depression (light gray) annotations at one of the keyhole collapse occurrences of Exp. A. The annotations are direct output of the trained U-Net.

For each peak identified in APS experiments A, B, C, D, we compared the x-ray image at the according time interval to see if a typical keyhole collapse cycle occurs.
Since the TED signals have different sensitivity on APS experiments and PART experiment due to their difference in substrates, the model's parameter was set to (H, M) = (0.03, 0.02)

# 3.Result

The results of this study demonstrate the effectiveness of the TED-based model in predicting pore formation during the LPBF process. By comparing the identified TED peaks with CT-scanned pore counts across eight samples, we validated the accuracy of the method, achieving a high correlation between TED signal peaks and actual pore counts.

## 3.1 Pores Count Prediction across 8 samples

The model with default parameter (H, M) = (0.335,0.2) was applied to 8 Part samples, then the number of TED peaks identified in each sample was recorded then compared with the number of pores detected obtained by CT scanning. As figure xxx shows, a strong correlation of R^2=0.94 was observed between the peak counts and the number of CT-detected pores.

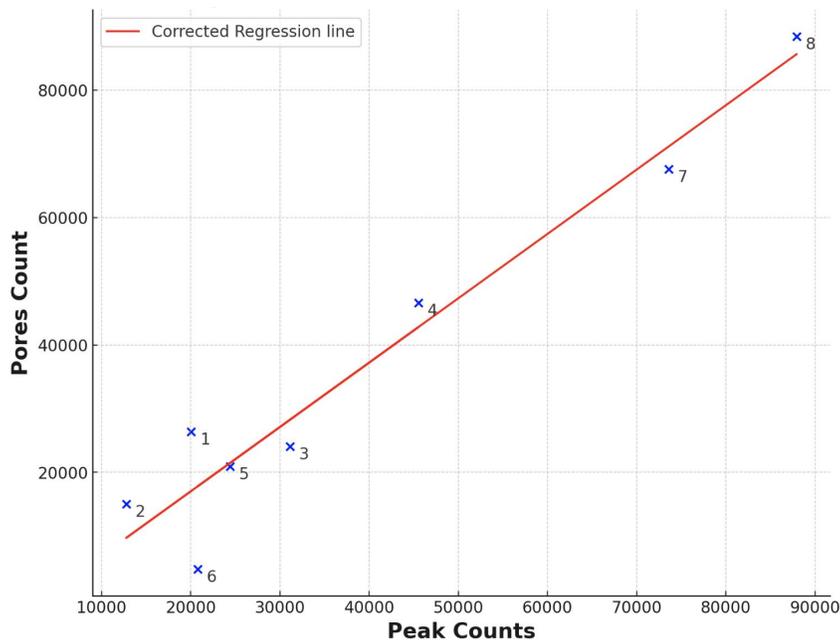

Fig xxx: Regression analysis between TED peaks count and pores count across 8 samples

Although the peaks identified by the TED have strong correlation with pores count, the relationship between TED peaks, keyhole collapse and porosity remain unclear: keyhole may collapse may happen during different time intervals, with TED peak patterns varying from case to case, and each collapse potentially generating multiple pores of varying sizes. Hence, we conducted a sensitivity analysis using different parameter combinations (H, M) to assess the model's performance in terms of R² and the deviation in the number of Thermal Energy Density (TED) peaks and pores.

First, a 3x3 sensitivity test is designed for (H, M) with a center point of (H, M) = (0.335, 0.2), using step sizes of 0.055 for H and 0.05 for M. The highest R2 value of 0.95 is achieved when H = 0.39 and M =0.25. Similar tests are also conducted for the absolute and normalized values of total TED counts across 8 samples. With 293868 pores in all 8 samples, the normalized TED peaks count is obtained by dividing each cell's count by 293868. The result of parameters sensitivity test is shown in figure xxx.

(a)
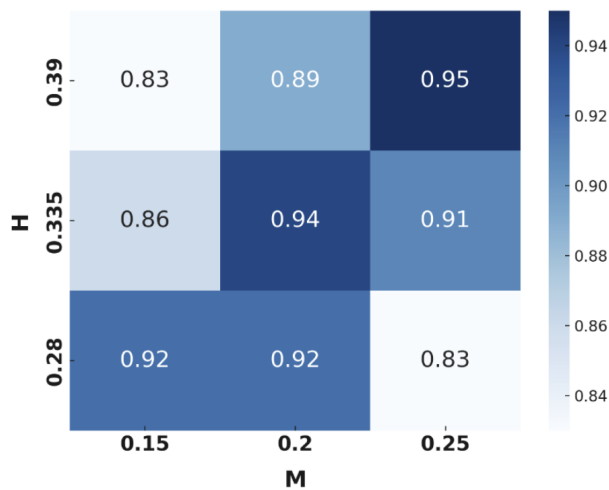

(b)
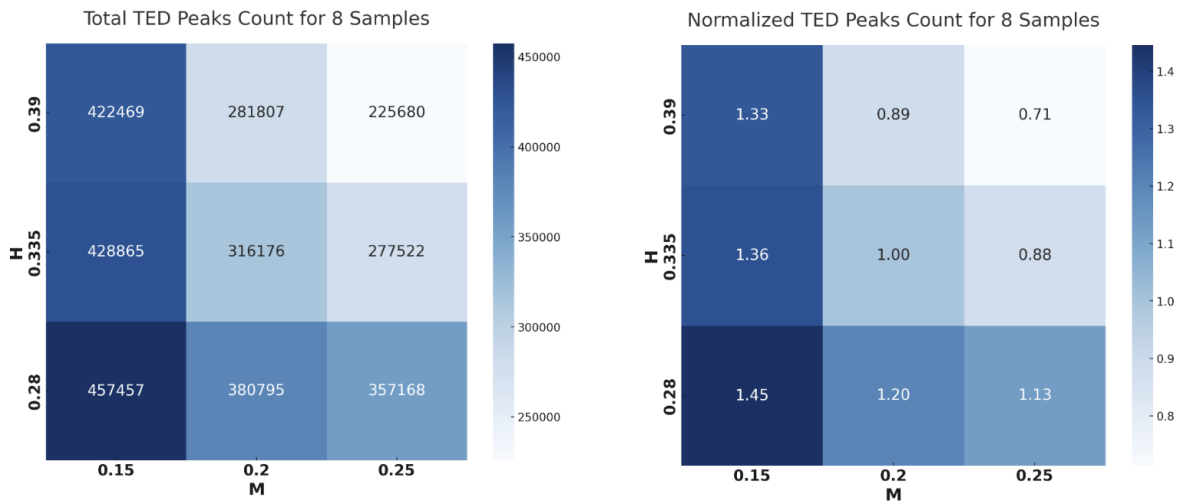
(b)

Figure xxx: sensitivity test result of the TED model. 9 parameters set (H, M) around the default value (H, M) = (0.335, 0.2) were tested. (a) shows the R^2 across 8 samples. (b) shows the total TED peaks count and normalized peaks count(right)

According to the sensitivity test, the default parameters (H, M) = (0.335,0.2) achieve the best pores count prediction in terms of both R^2 and absolute values. But this does not mean each TED (keyhole collapse) identified generates one pore. Such high correlation may be explained by a more advanced probability model.

## 3.2 Layer-wise Prediction

The layer-wise distribution of TED peak counts predicted by the model match well with the layer-wise distribution of pore counts across all eight samples. Among these, sample 1,2,3,4,5,8 are printed with regular parameters, while samples 6 and 7 are exceptional cases: sample 6 was made with optimized processing parameters that reduced pore count in the bulk section, while sample 7 was made with Lack of Fusion pores.

### 3.2.1 Layer-wise prediction for samples made by regular processing parameter

For selected samples from regular process parameter, the layer-wise TED peaks identified by the model and pores count are plotted below.

**(a)**

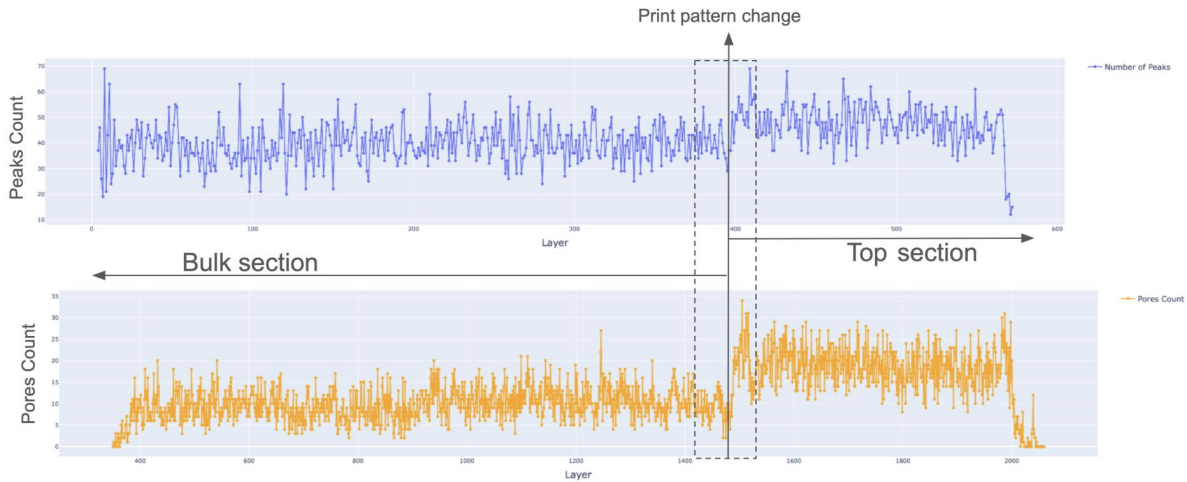

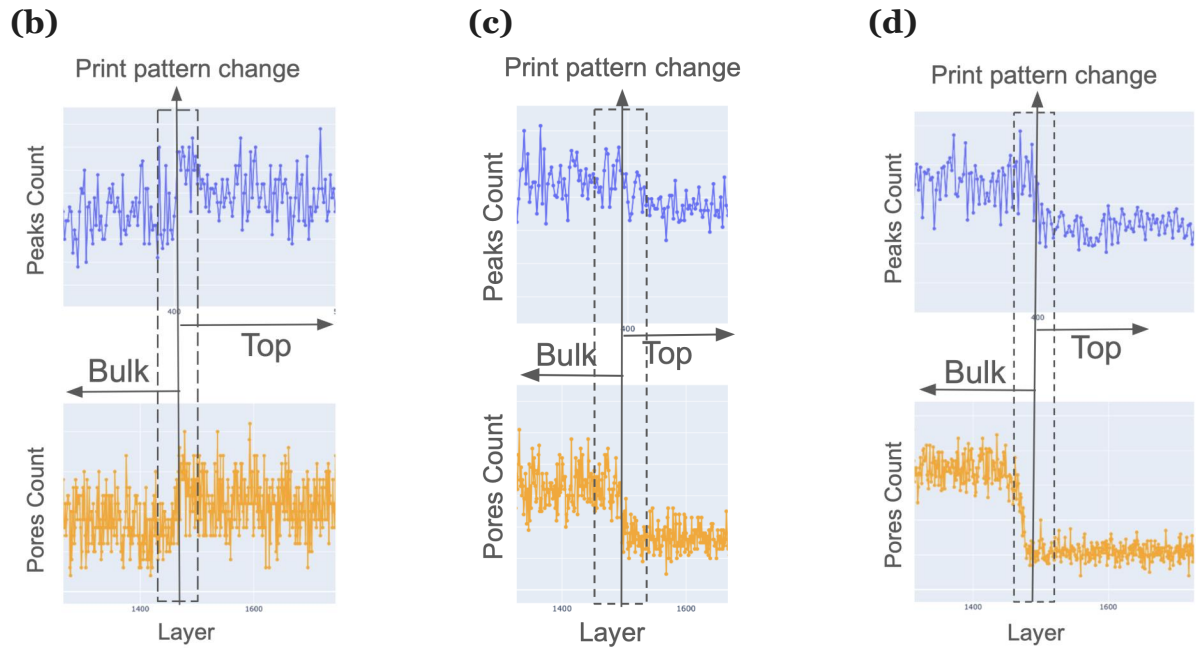

Figure xxx: Comparison of layer-wise TED peak counts and CT-scanned pore counts for samples printed with regular process parameter. The TED peaks distribution and pores distribution are plotted in blue and orange respectively. In both the bulk and top sections, the six regular samples exhibit an even distribution of peaks and pores for a single printing pattern. The model effectively captures changes in pore distribution across different printing patterns (bulk or top). (a) illustrates the distribution of peaks and pores across all layers in sample 5 (layers 1–573), while (b), (c), and (d) correspond to samples 2, 4, and 8, respectively, focusing on regions around layer 400.

For six regular samples, the peaks identified by the model successfully predict the layer-wise distribution of pores: for all 6 samples, the peaks' layer-wise fluctuation match with that of pores in a single printing pattern; the peaks also capture the porosity change due to printing patterns change, which happens in a single printing layer. This suggests the model can predict layer-wise pores distribution during LPBF.

### 3.2.2 Layer-wise prediction for samples made by special processing parameter

Two samples with special process parameters were designed in the experiment:

- Sample 7 was made intentionally lack of fusion pores, and the previous research about TED peaks only shows that TED peaks match with keyhole collapses, instead of lack-of-fusion pores.
- Sample 6 was made by an optimized parameter to reduce pores. While TED peaks monitor the real-time keyhole collapse and potential pores formation, some cross-layer pores healing mechanisms may have occurred during the printing.

Comparison of layer-wise TED peaks and pores distribution for sample 6 and 7 is shown below:

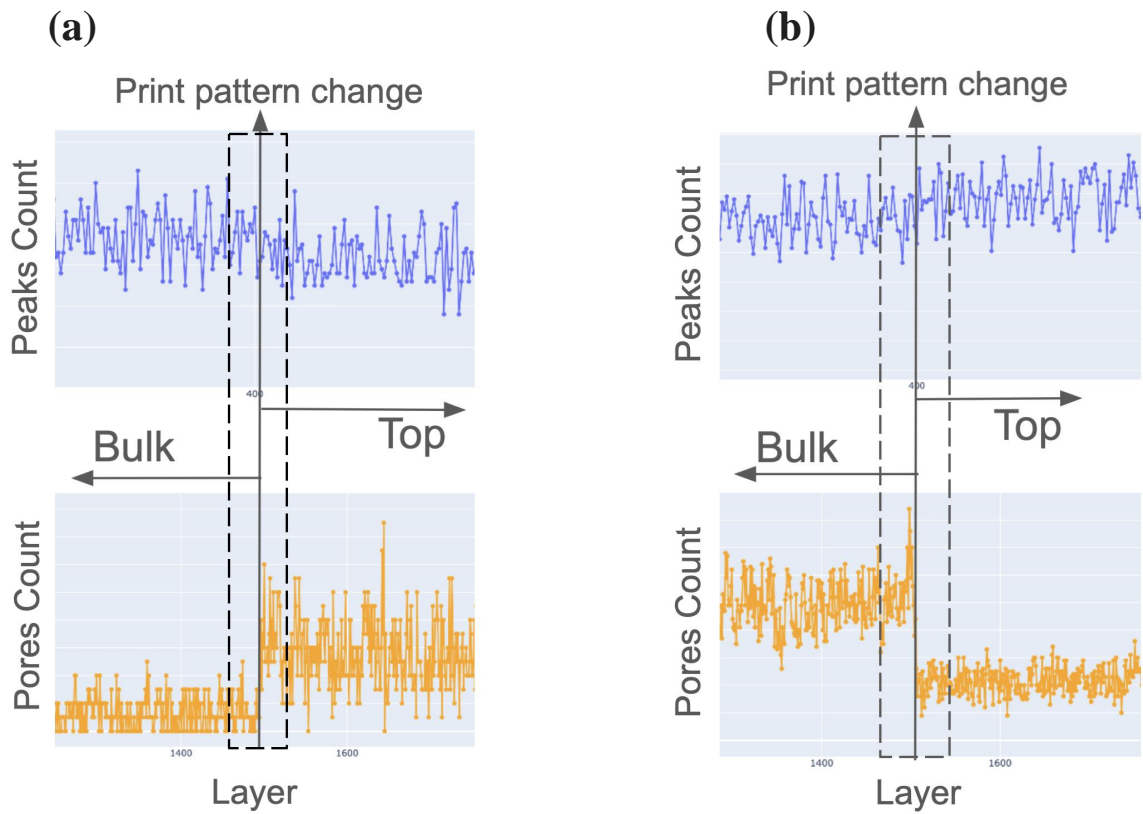

Fig xxx: Comparison of layer-wise TED peak counts and CT-scanned pore counts for samples printed with special process parameter. The TED peaks distribution and pores distribution are plotted in blue and orange, respectively. Plots (a), (b) correspond to sample 6, 7 respectively.

Sample 6 and 7 exhibit an even distribution of peaks and pores for a single printing pattern. However, the peaks fail to predict the pores distribution before and after the transition point. While such discrepancy helped explain why sample 6 was an "outlier" in the while part prediction, we don't have a good sense why the total pores and peaks count are very close in sample 7 but the layer-wise trended doesn't match. For sample 6, its bulk part has a very sparse pores distribution. Considering that TED values record the real-time keyhole collapse, we guess some cross- layer pores healing mechanism may occur.

## 3.3 Compare the identified TED peaks with X-Ray ground truth

The validity of the TED peaks identification method, i,e, if it catches a keyhole collapse, was examined by observing the high-speed X-Ray images during the corresponding time interval to see whether a "keyhole collapse" happened or not. Such "ground truth test" was conducted for APS experiments since the PART experiments were not equipped high-speed X-Ray. Since the TED signals have different sensitivity on APS experiments and PART experiment due to their difference in substrates, the model's parameter was set to (H, M) = (0.03, 0.02).

After analyzing the X-ray images corresponding to the intervals of the identified TED peaks, we observed that the TED peaks perfectly align with keyhole collapses. As previously mentioned, a typical keyhole collapse includes four stages: Stable >> Expansion >> Collapse >> Recovery. Figure xxx presents the X-ray ground truth from experiment B, illustrating these stages. Stages "Stable", "Expansion", "Collapse", "Recovery", matched with label "a", "b", "c", "d", could be clearly see in a consecutive time interval. The TED "Peak" interval roughly corresponds to the "Expansion" and "Collapse" interval, but the image of "Stable" and "Recovery" stages were add to illustrate the complete keyhole collapse cycle.

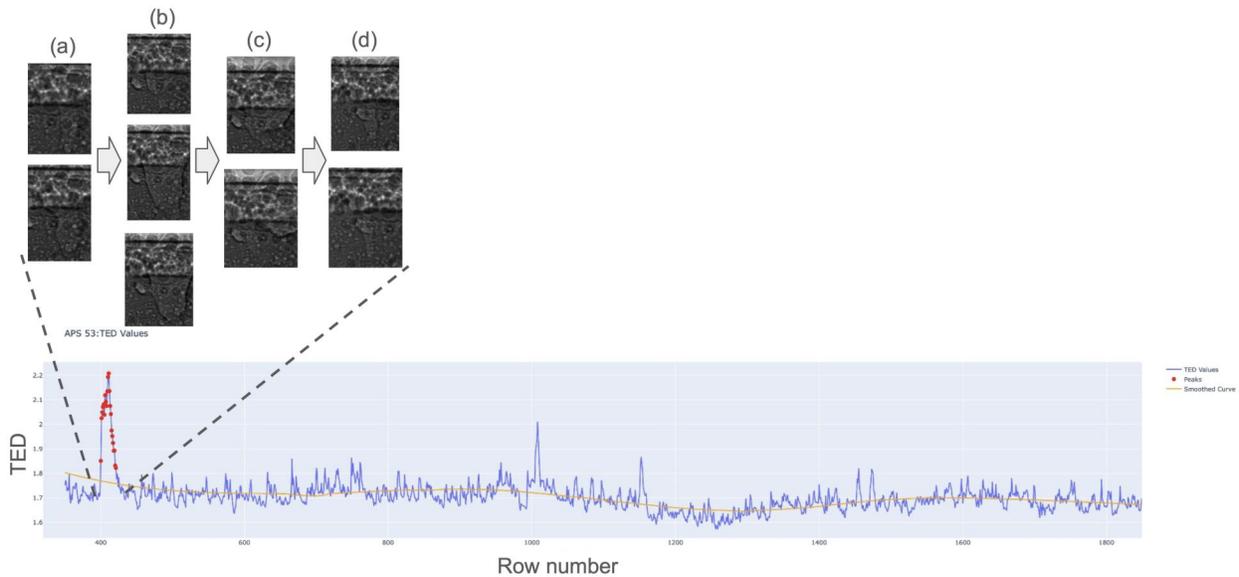

Fig xxx: X-ray images of the TED peak interval identified in experiment B, captured over consecutive time intervals, reveal all four stages of a typical keyhole collapse within the identified peak interval. (a), (b), (c), and (d) correspond to the "stable," "expansion," "collapse," and "recovery" stages, respectively.

In four APS experiments, six TED peaks were identified. Typical keyhole collapses, like those shown in Figure xxx, were observed in five of them. The only exception happens in experiment D, the powder thickness was twice that of the other experiments, preventing the laser from penetrating the powder. Consequently, the keyhole remained concealed within the powder, making it hard to determine whether a keyhole collapse occurred. Therefore, for TED peaks with clear X-ray image comparisons, they have a 100% prediction accuracy for keyhole collapse. This explains the high correlation between the TED peaks Count and Pores Count.

# 4.Discussion

## 4.1 Theoretical basis of the model

To the best of our knowledge, there doesn't exist a complete theory that fully explains the instability of keyholes during the LPBF process. However, we propose a possible explanation for why TED signals generate peaks during keyhole collapse. In the model, these TED peaks are defined as "consecutive intervals where TED values are consistently high." We hypothesize that when the laser strikes the powder during LPBF, energy is radiated from the laser-scanned spot in relation to the energy density at that spot. During the expansion phase, the keyhole (gas phase) enlarges, which indicates an increase in temperature in the keyhole region, linking to a rise in thermal energy density (TED). During the collapse phase, the keyhole contracts, indicating a drop in temperature and explaining the reduction in TED values. This hypothesis may explain how TED peaks correlate with keyhole collapse events.

## 4.2 Correlation Between TED Peaks and Pores

In this study, we observed a strong correlation between TED peaks and the number of pores formed, suggesting they are linked. While TED peaks are highly effective in predicting keyhole collapse events, there is no direct evidence suggesting that each TED peak corresponds to a pore. Instead, such a strong correlation appears to be caused by a series of probabilistic events. Given the randomness in pore formation following a keyhole collapse, a single collapse might result in more than one pore, or potentially no pore at all. Additionally, since 3D printing is a layer-by-layer process, the laser in subsequent layers can influence the distribution of pores formed in earlier layers. The method we introduce here—due to its low cost, simplicity, accuracy, and real-time capabilities—has the potential to be applied as an industrial quality inspection method for metal 3D printing. However, the underlying mechanism linking keyhole collapse to pore formation requires further investigation.

## 4.3 The potential of TED Peaks on Pore Size Distribution and Mechanical Properties

By analyzing TED peaks alongside high-speed X-ray imaging, we observed that the size of each TED peak varies depending on the keyhole collapse event. Larger peaks tend to correlate with larger keyholes during the expansion phase, which may lead to the formation of larger pores during the collapse stage since it is harder for the solidified material to fill out the gas region. This finding suggests that TED peaks not only predict the number of pores but also have the potential to predict pores size distribution

Previous studies have shown that pore size distribution significantly impacts various mechanical properties such as compressive strength, elastic modulus, fatigue resistance, and flexural strength.[12] Based on this, we hypothesize that TED values, when integrated into more advanced predictive models, could offer insights into the mechanical properties of 3D-printed parts. This research hints at a future where metal additive manufacturing may enter a "predictable" era, where critical mechanical properties can be forecasted in real time during the printing process.

## 4.4 Potential Extension Projects and Applications of the Pores Detection Method

In metal 3D printing, monitoring the condition of parts in real-time has been a big challenge. However, this research brings this vision closer to reality. During the execution of the program, we discovered that processing the entire sample—consisting of 573 layers with millions of TED datapoints recorded—through the TED peaks identification procedure takes approximately one minute, which is negligible compared to the overall printing time. This suggests that if the TED signal is input into the program in real-time during printing, the program can immediately provide the distribution of pores throughout the printing process. This allows operators to understand the sample's pore distribution for any given time interval and make informed decisions accordingly.

Also, researches have shown that process parameters can be automatically adjusted during printing to reduce pore formation [13]. The precise prediction of pores for each time interval based on TED signals could offer a more accurate reference for automated parameter optimization.

Moreover, in this study, we used AlSiMg material. But keyhole collapse, which leads to pore formation, is a common phenomenon in metal 3D printing. Therefore, we believe that the method used to detect keyhole collapse in this study can also be applied to other metal 3D printing materials, such as steel and titanium alloys. However, due to variations in the absorption rates of different materials when exposed to laser energy, the model parameters would need further adjusted.

# Conclusion

In conclusion, this study presents a novel, cost-effective, and real-time pores detection method utilizing Thermal Energy Density (TED) signals during the Laser Powder Bed Fusion (LPBF) process. The method demonstrates a strong correlation between TED signal peaks and keyhole collapse, offering an accurate prediction of exact pores count. Unlike traditional detection methods that are time-consuming and rely on costly equipment, our approach is scalable for industrial applications and capable of capturing layer-wise pore distribution trends. Future work may explore deeper relationships between TED signal characteristics and the mechanical properties of printed parts, potentially allowing for real-time process optimization to improve product quality across different materials and geometries.